\documentclass[aps,prl,twocolumn,groupedaddress,amsmath,amssymb,superscriptaddress]{revtex4-2}

\usepackage{graphicx}
\usepackage{bm}
\usepackage[none]{hyphenat}
\usepackage[draft]{changes}

\begin{document}

\title{Intrinsic Moiré Higher-Order Topology Beyond Effective Moiré Lattice Models}

\author{Xianliang Zhou}
\thanks{These authors contributed equally.}
\affiliation{Key Laboratory of Artificial Structures and Quantum Control (Ministry of Education), School of Physics and Astronomy, Shanghai Jiao Tong University, Shanghai 200240, China}
\affiliation{Tsung-Dao Lee Institute, Shanghai Jiao Tong University, Shanghai, 200240, China}
\affiliation{Institute of Physics, Chinese Academy of Sciences, Beijing 100190, People’s Republic of China}

\author{Yifan Gao}
\thanks{These authors contributed equally.}
\affiliation{Department of Physics, Southern University of Science and Technology, Shenzhen, Guangdong 518055, China}

\author{Laiyuan Su}
\affiliation{Department of Physics, Southern University of Science and Technology, Shenzhen, Guangdong 518055, China}

\author{Z.~F. Wang}
\affiliation{Hefei National Research Center for Physical Sciences at the Microscale, CAS Key Laboratory of Strongly-Coupled Quantum Matter Physics, Department of Physics, University of Science and Technology of China, Hefei, Anhui 230026, China}
\affiliation{Hefei National Laboratory, University of Science and Technology of China, Hefei, Anhui 230088, China}

\author{Li Huang}
\affiliation{Department of Physics, Southern University of Science and Technology, Shenzhen, Guangdong 518055, China}
\affiliation{Quantum Science Center of Guangdong-HongKong-Macao Greater Bay Area (Guangdong), Shenzhen 518045, China}

\author{Angel Rubio}
\affiliation{Max Planck Institute for the Structure and Dynamics of Matter, Center for Free Electron Laser Science, 22761 Hamburg, Germany}
\affiliation{Initiative for Computational Catalysis (ICC) and Center for Computational Quantum Physics (CCQ), Simons Foundation Flatiron Institute, New York, NY 10010}

\author{Zhiwen Shi}
\email{zwshi@sjtu.edu.cn}
\affiliation{Key Laboratory of Artificial Structures and Quantum Control (Ministry of Education), School of Physics and Astronomy, Shanghai Jiao Tong University, Shanghai 200240, China}
\affiliation{Tsung-Dao Lee Institute, Shanghai Jiao Tong University, Shanghai, 200240, China}
\affiliation{Collaborative Innovation Center of Advanced Microstructures, Nanjing University, Nanjing 210093, China}

\author{Lede Xian}
\email{ldxian@tias.ac.cn}
\affiliation{Tsientang Institute for Advanced Study, Zhejiang 310024, China}
\affiliation{Songshan Lake Materials Laboratory, Dongguan, Guangdong 523808, China}
\affiliation{Max Planck Institute for the Structure and Dynamics of Matter, Center for Free Electron Laser Science, 22761 Hamburg, Germany}


\begin{abstract}
Moiré superlattices provide a compelling platform for exploring exotic correlated physics. Electronic interference within these systems often results in flat bands with localized electrons, which are typically described by effective moiré lattice models. While conventional models treat moiré sites as indivisible, analogous to atoms in a crystal, this picture overlooks a crucial distinction: unlike a true atom, a moiré site is composed of tens to thousands of atoms and is therefore spatially divisible. Here, we introduce a universal mechanism rooted in this spatial divisibility to create topological boundary states in moiré materials. Through tight-binding and density functional theory calculations, we demonstrate that cutting a moiré site with a physical boundary induces bulk topological polarization, generating robust boundary states with fractional charges. We further show that when the net edge polarization is canceled, this mechanism drives the system into an intrinsic moiré higher-order topological insulator (mHOTI) phase. As a concrete realization, we predict that twisted bilayer tungsten disulfide (WS$_2$) is a robust mHOTI with  experimentally detectable corner states when its boundaries cut through moiré hole sites. Our findings generalize the theoretical framework of moiré higher-order topology, highlight the critical role of edge terminations, and suggest new opportunities for realizing correlated HOTIs and higher-order superconductivity in moiré platforms.
\end{abstract}
\maketitle

Moiré superlattices provide a compelling platform for exploring exotic phenomena, such as Mott insulation \cite{RN1,RN2,RN3,RN4,RN5,RN6,RN7}, unconventional superconductivity \cite{RN8,RN9,RN10,RN11,RN441}, and the quantum anomalous Hall effect \cite{RN12,RN13,RN14,RN15,RN16,RN17}. The rich phenomenology arises from electronic interference, which generates nearly flat bands with quenched kinetic energy. The electronic states in these flat bands are spatially localized, forming arrays of artificial ``moiré atoms" that can be described by effective lattice models on moiré lattice sites \cite{EF1,EF2,EF3,quantum_simulator}. Such effective models accurately capture the low-energy physics of moiré systems and have successfully explained a range of correlated phenomena, including Mott insulators \cite{RN18,RN19}, Wigner crystals \cite{RN20,RN21,RN22,RN23}, and Wigner molecules \cite{RN24,RN25,RN445}.

\begin{figure}
\includegraphics[width=8.6cm]{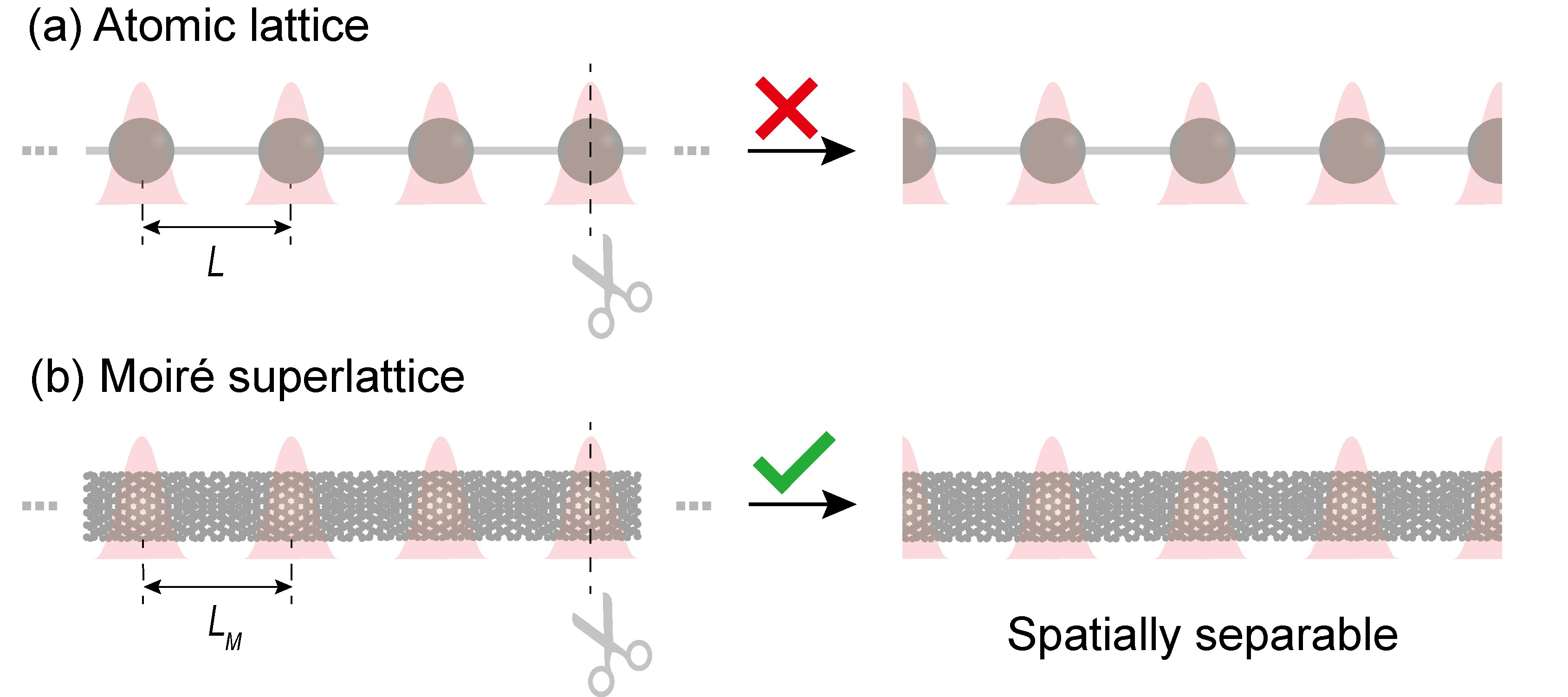}
\centering
\caption{
(a) In atomic lattices, each site corresponds to a single atom, making it impossible to "cut" the site. (b) In moiré superlattices, each moiré site corresponds to a spatially extended electron orbital spanning many atoms, allowing moiré lattice sites to be, in principle spatially divisible.
}\label{fig:fig1}
\end{figure}

Conventional effective models treat moiré sites as indivisible, much like atoms in a crystal. This picture, however, overlooks a crucial distinction: unlike true atoms, moiré site consists of tens to thousands of atoms and can therefore be spatially divided into fractions [Fig.~\ref{fig:fig1}]. This spatial divisibility introduces a new degree of freedom—the ability to terminate the moiré lattice by “cutting” through a moiré site with a physical boundary. Such an operation is impossible in atomic crystals and fundamentally alters the boundary physics, providing a new route to engineering moiré topology.

In this Letter, we introduce a universal and versatile mechanism for creating topological boundary states in moiré materials, rooted in the spatial divisibility of moiré sites. Using tight-binding (TB) and density functional theory (DFT) calculations on one- and two-dimensional moiré systems, we demonstrate that cutting a moiré site with a physical boundary induces a robust edge polarization, that generates topological boundary states. This mechanism, independent of crystalline symmetry, results in an intrinsic moiré higher-order topological insulator (mHOTI) phase when the net edge polarization is canceled (e.g., through a layer degeneracy). As a concrete realization, we predict that twisted bilayer tungsten disulfide (tbWS$_2$) is a mHOTI, hosting experimentally detectable HOTI corner states. Our work establishes cutting of moiré site as a new design principle for moiré topology and opens new avenues for exploring correlated HOTIs and higher-order superconductivity.

We first investigate the consequences of cutting the moiré site in a one-dimensional (1D) system. To capture the essential physics, we consider a 1D atomic chain subjected to a slowly varying moiré potential [Fig. 2(a)], governed by the TB Hamiltonian:
\begin{equation}
H_{1 D}=-\sum_it c_{i+1}^{\dagger} c_i+\sum_iV_i c_i^{\dagger} c_i+\text {h.c.},
\label{eq1}
\end{equation}
where $i$ represents a 1D lattice site, $c_i^{\dagger}$ ($c_i$) is the creation (annihilation) operator for an electron at site $i$, and $t$ is the hopping amplitude. While the effective moiré potential $V_i$ in real materials can be complex \cite{RN31,RN32}, for simplicity we adopt a harmonic moiré potential with inversion symmetry:
\begin{equation}
V_i=v \cos \left(\frac{2 \pi i}{L_M}-\frac{\pi}{L_M}\right),
\label{eq2}
\end{equation}
where $v$ is the amplitude of the moiré potential, and $L_M$ is the moiré period.

\begin{figure}
\includegraphics[width=8.6cm]{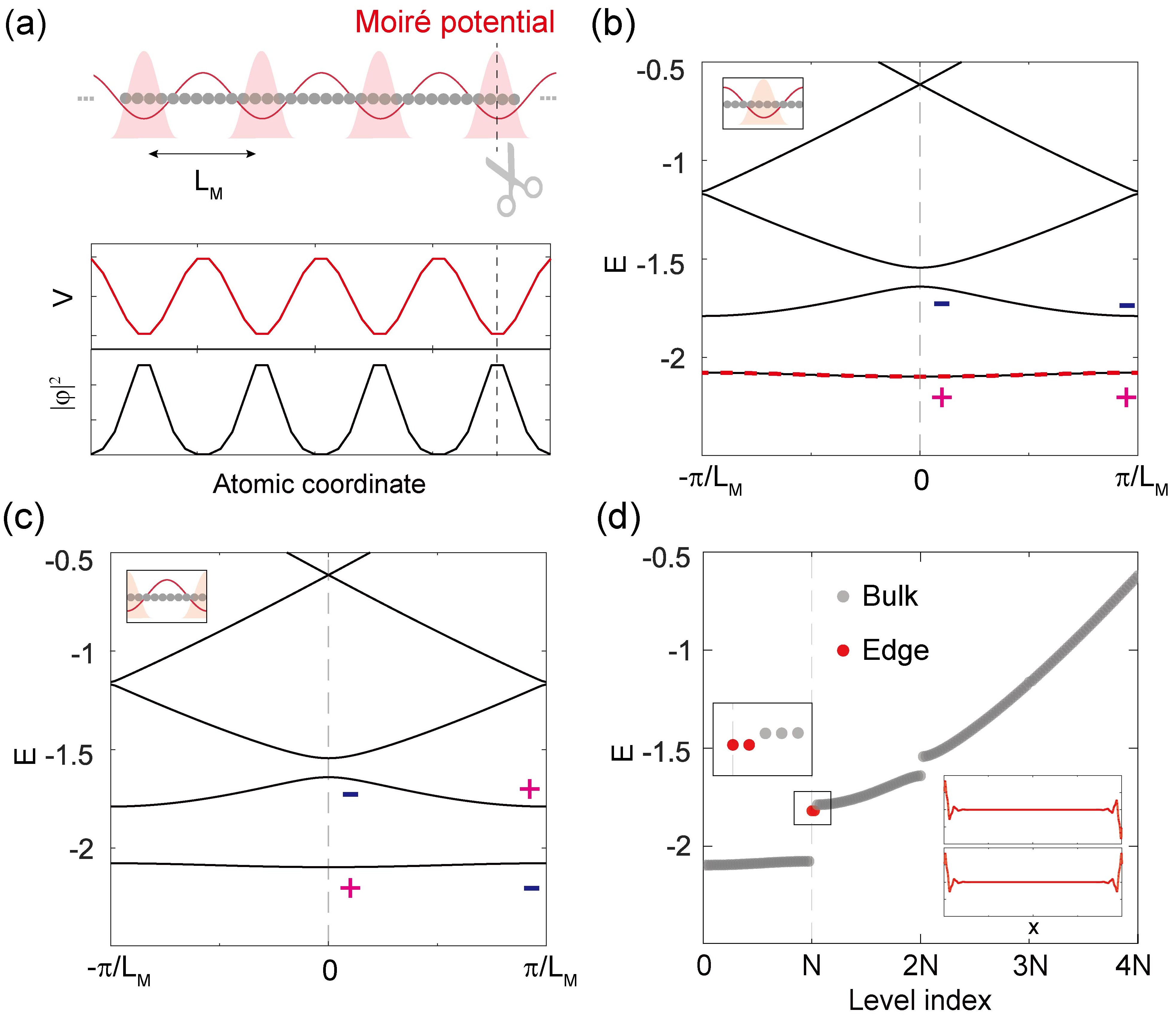}
\centering
\caption{\label{fig2}Emergent topological boundary states in a 1D moiré superlattice. (a) Schematic of the simplified 1D moiré superlattice model. Lower panel shows the moiré potential and the charge density of the bottom band. (b) Moiré band structure with a conventional unit cell (inset) for $v/t=0.3$ and $L_M=10$. The red dashed line represents a fit of effective lattice model. Parity eigenvalues at high-symmetry points are labeled by “$\pm$”. (c) Moiré band structure with an unconventional unit cell (inset) for $v/t=-0.3$. (d) Open-boundary energy spectrum of (c), with $N=40$ unit cells. The x-axis represents the total number of electrons, where N, 2N, 3N correspond to the full filling of the first, second, and third bands in (c), respectively. Inset shows the wavefunctions of the two topological boundary states at full filling of the first band (dashed line).}
\end{figure}

The calculated band structure of our 1D model [Fig. 2(b)] reveals prominent gaps, isolating the two lowest bands. The charge density of the lowest band is strongly localized at the moiré potential minima [see lower panel, Fig. 2(a)], forming an effective moiré lattice. The low-energy physics of the 1D moiré system is accurately captured by a conventional effective lattice Hamiltonian:
\begin{equation}
H= \sum_{R, R^{\prime}} t\left(R-R^{\prime}\right) c_{R }^{\dagger} c_{R^{\prime}},
\label{effective}
\end{equation}
where $R$ denotes the moiré lattice sites. As confirmed in Fig. 2(b), this model perfectly reproduces the dispersion of the lowest moiré band.

Because moiré sites are spatially extended over many atoms [Fig. 2(a)], we can choose an unconventional unit cell whose boundary intersects a potential minimum  [inset, Fig. 2(c)]. This choice of boundary determines the system's termination under open boundary conditions, where the boundary cuts the moiré sites at the open edges. This action creates a charge imbalance at full band filling known as a filling anomaly \cite{PhysRevB.99.245151}, which can be characterized by the bulk electric polarization (Zak phase) \cite{RN35,RN33}:
\begin{equation}
P_n=\frac{1}{2\pi}\int_{\mathrm{BZ}} d k A_n(k),
\label{eq3}
\end{equation}
where $A_n$ is the Berry connection of band $n$. Under inversion symmetry, $P_n$ is constrained to quantized values determined by parity eigenvalues $\eta_n (k)$ at high-symmetry points $\Gamma$ and $X$ \cite{RN34}:
\begin{equation}
(-1)^{2P_n }=\frac{\eta_n(\mathrm{X})}{\eta_n(\Gamma)}.
\label{eq4}
\end{equation}

For the two lowest bands, we find the parity eigenvalues at $\Gamma$ and $X$ are distinct, yielding a nontrivial polarization $P_1=P_2=1/2$. This result, which contrasts sharply with the trivial polarization ($P=0$) of the system with a conventional unit cell [Fig. 2(b)], provides definitive evidence that such unconventional choice of unit cell induces a band inversion, transforming the system into a 1D topological insulator.

To confirm this topology, we calculate the energy spectrum for the system under open boundary conditions with N unit cells, as shown in Fig. 2(d). The bulk states (gray) exhibit the same energy range and gaps as the periodic system. At the full filling of the lowest band (N electrons filling), the nontrivial polarization $P_1=1/2$ manifests a filling anomaly, with emergence of two degenerate, inversion-protected boundary states (red). These states are localized at opposite ends of the chain, with each carrying a fractional charge of e/2 [inset, Fig. 2(d)]. In contrast, when both bands are filled (2N electrons filling), the total polarization becomes trivial and the boundary states vanish, because $P_{\text{tot}}=\sum_n P_n= P_1+P_2=0 \left(\bmod~1\right)$, where the summation is over occupied bands. Notably, the lowest two bands in Fig. 2(c) are topologically equivalent to the Su–Schrieffer–Heeger (SSH) model \cite{PhysRevLett.42.1698}, with the system's topological properties similarly dependent on boundary termination. However, the nearly continuous nature of moiré cutting (see Supplemental Material \cite{supplementary}) offers a new perspective on the topological-to-trivial phase transition, a feature not present in the SSH model.

To further support our claim, we examine two additional 1D systems: a uniaxially strained bilayer graphene nanoribbon and a DFT-simulated Beryllium-Hydrogen atomic chain (see Supplemental Material \cite{supplementary}). Both systems host robust topological edge states, but only when their boundaries cut the moiré localization sites. This finding establishes the cutting of moiré lattice site as a broadly applicable design principle for engineering moiré topology.

Our analysis readily extends to two-dimensional (2D) moiré systems. We consider a model system consisting of a 2D square lattice subjected to a slowly varying moiré potential [Fig. 3(a)], described by the following TB Hamiltonian:
\begin{equation}
\begin{split}
    H_{2 D}
    =&- \sum_{i, j}\mathrm{t}\left(c_{i+1, j}^{\dagger} c_{i, j}+c_{i, \mathrm{j}+1}^{\dagger} c_{i, j}\right)\\
    &+\sum_{i, j} V_{i, j} c_{i, j}^{\dagger} c_{i, j}+\text {h.c.},
\end{split}
\label{eq5}
\end{equation}
where $(i,j)$ denotes a 2D lattice site. For simplicity, we consider a harmonic moiré potential $V_i$ that respects both inversion and $C_4$ rotational symmetry, defined as:
\begin{equation}
V_{i, j}=v \cos \left(\frac{2 \pi i}{L_M}-\frac{\pi}{L_M}\right)+v \cos \left(\frac{2 \pi j}{L_M}-\frac{\pi}{L_M}\right).
\label{eq6}
\end{equation}

The low-energy bands of the 2D model is captured by an effective model on a moiré square lattice, detailed in the Supplemental Material \cite{supplementary}. To investigate the consequences of this moiré lattice's spatial divisibility, we calculate the band structure for a unit cell whose boundaries intersect a moiré site [Fig. 3(a)]. Simiar to the 1D case, this configuration creates a filling anomaly at full band filling under open boundary conditions, which can be characterized by the 2D electric polarization $\mathbf{P}_n$. This polarization is determined by the parity eigenvalues at the high-symmetry points $\Gamma$ and $X(Y)$ \cite{RN35,RN34}:
\begin{equation}
(-1)^{ 2P_n^i}=\frac{\eta_n(\mathrm{X_i})}{\eta_n(\Gamma)},
\label{eq7}
\end{equation}
where $i=x,y$. For the bottom four bands, the parity eigenvalues are distinct between the $\Gamma$ and $X(Y)$ points [Fig.~3(b)], yielding a nontrivial polarization $\mathbf{P}=(1/2,1/2)$. This result demonstrates that such choice of unit cell induces band inversion, thereby opening a topological gap in the 2D system.

\begin{figure}
\includegraphics[width=8.6cm]{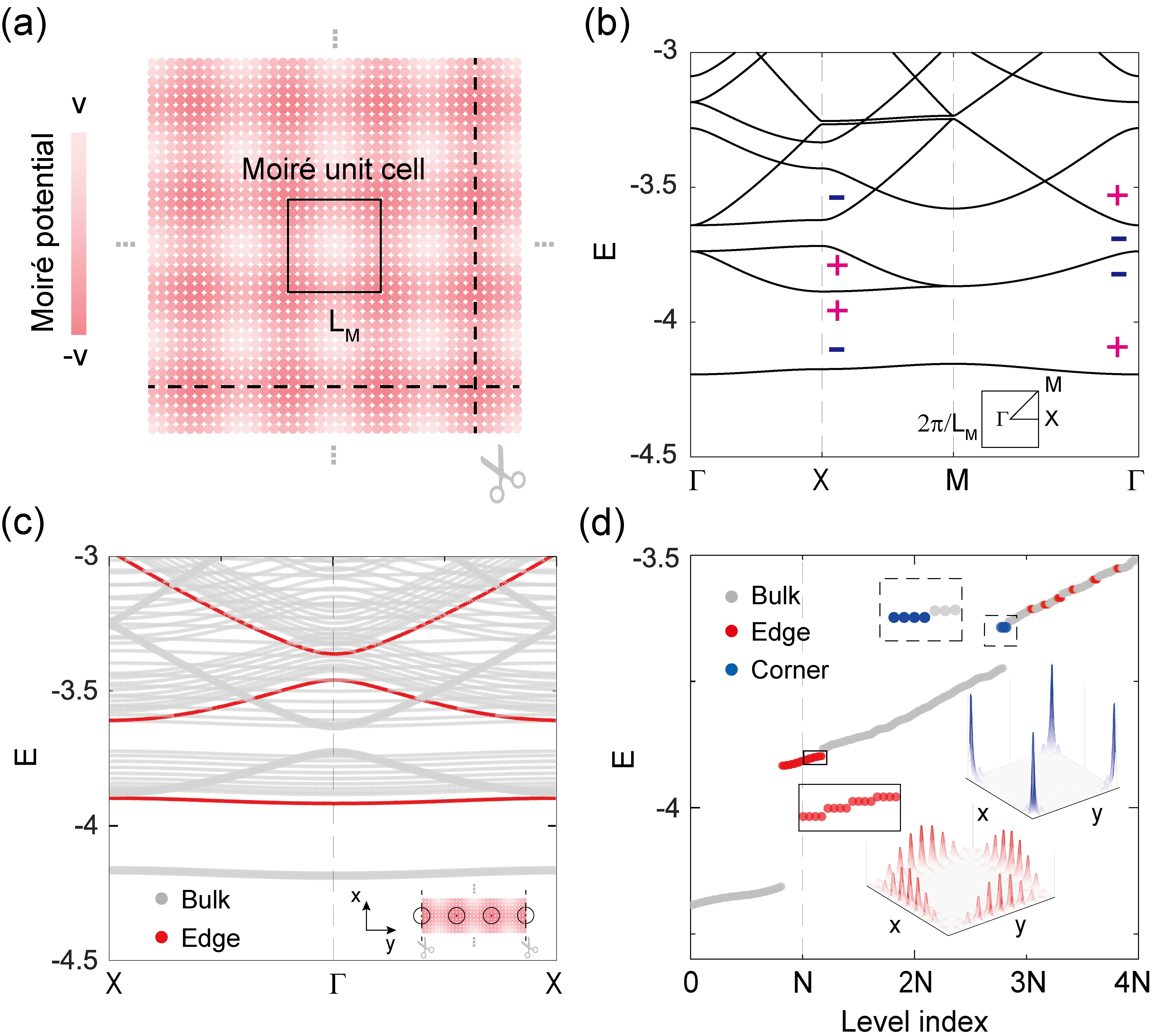}
\centering
\caption{\label{fig3}Emergent topological boundary states in a 2D moiré superlattice. (a) Schematic of the square lattice model with a moiré potential. The solid box  marks the moiré unit cell. (b) Moiré band structure for $v/t=-0.3$ and $L_M=10$. Inset shows the moiré Brillouin zone. Parity eigenvalues at high-symmetry points are labeled by “$\pm$”. (c) Ribbon spectrum of the band structure in (b), finite along the $y$-direction ($N_y=10$). Inset shows the ribbon geometry. (d) Energy spectrum of the band structure in (b), of a finite flake with $N=N_x\times N_y=10\times 10=100$ unit cells. Inset shows the charge density of edge states (bottom red) and corner states (top blue).}
\end{figure}

The ribbon spectrum of 2D model [Fig. 3(c)], calculated for a finite geometry along the $y$-direction, reveals topological edge bands (red) emerging from the bulk's filling anomaly. Their dispersion closely resembles that of the 1D moiré model [Fig. 2(c)]. Under full open-boundary conditions, the energy spectrum [Fig. 3(d)] further displays four degenerate, $C_4$-protected topological corner states (blue). These states originate from the occupation of 1/4 moiré sites at open corners and are highly localized, carrying fractional charges of $e/4$ [inset, Fig. 3(d)]. The emergence of these edge and corner states is independent of the underlying crystal symmetry, and they are robust against symmetry-breaking perturbations. Crucially, all boundary states vanish when boundaries avoid the moiré sites (Supplementary Material \cite{supplementary}), confirming that the new concept proposed in this work of cutting moiré sites is the essential mechanism. It is noteworthy that this topology phase stems from a nontrivial 2D electric polarization, distinguishing it from conventional topological insulators \cite{PhysRevLett.118.076803}.

The edge polarization in 2D moiré systems can be eliminated through homobilayer stacking, which allows a mHOTI phase \cite{PhysRevB.96.245115} to emerge by cutting moiré sites in twisted homobilayer systems. We demonstrate this intrinsic moiré higher-order topology in tbWS$_2$ as an representative example [Fig. 4(a)], applying a $\Gamma$-valley valence bands effective Hamiltonian model \cite{doi:10.1073/pnas.2021826118}:

\begin{figure}
\includegraphics[width=8.6cm]{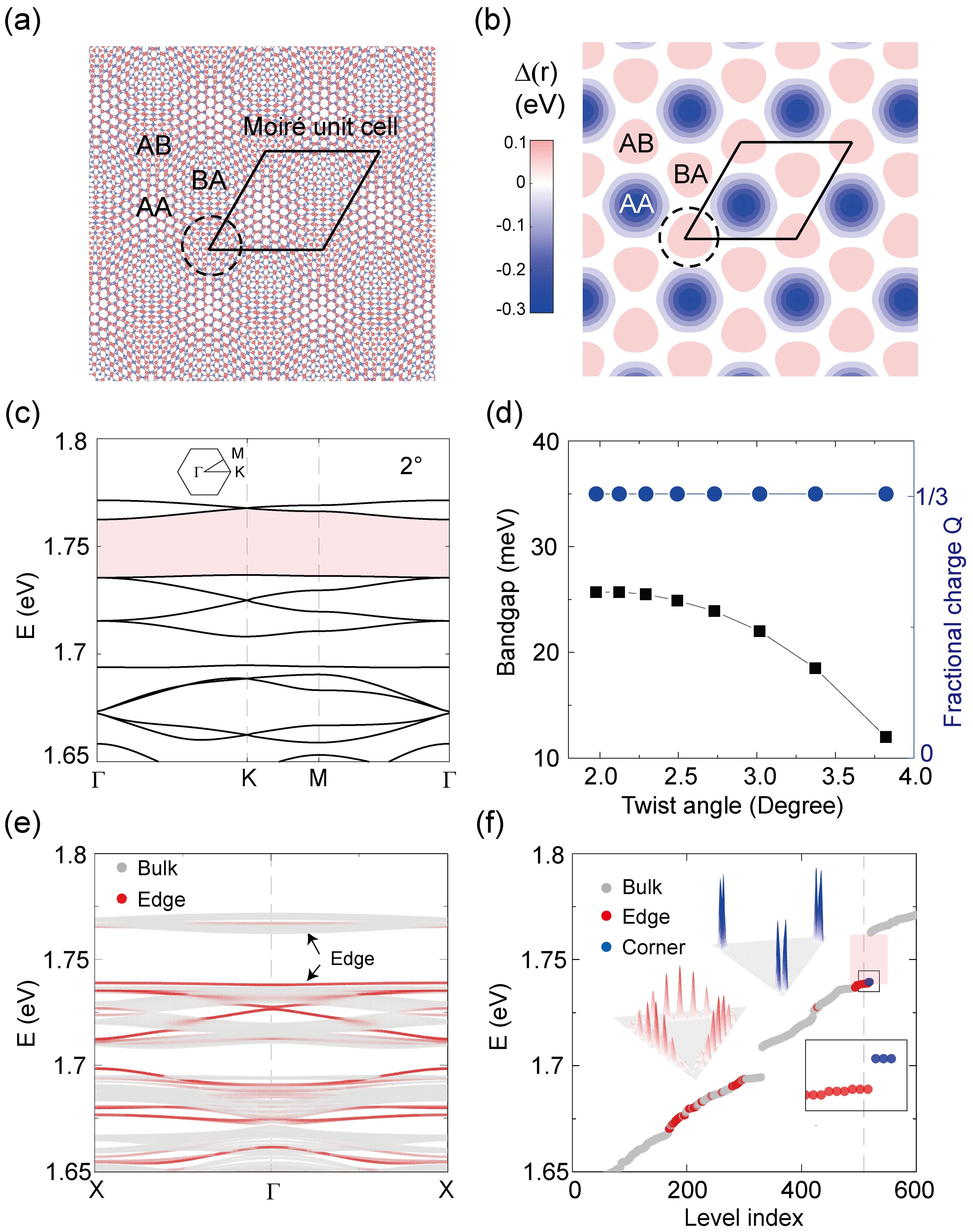}
\centering
\caption{\label{fig4} 2D mHOTI in tbWS$_2$. (a) Atomic structure and moiré unit cell of tbWS$_2$. (b) Moiré potential felt by holes at VBM in tbWS$_2$. (c) Moiré band structure for a twist angle of $\theta = 2^\circ$. Inset shows the Brillouin zone. (d) First Moiré band gap and fractional corner charge $Q$ as a function of twist angles. (e) Ribbon spectrum corresponding to the band structure in (c). (f) Energy spectrum under full open-boundary conditions for a $10\times10$ supercell. Inset shows the charge density of edge states and HOTI corner states. }
\end{figure}

\begin{equation}
H=-\frac{\hbar^2 k^2}{2 m^*}+\Delta(\mathbf{r})
\end{equation}
\begin{equation}
\Delta(\mathbf{r})=\sum_s \sum_{j=1}^6 V_s \exp \left(i \mathbf{g}_j^s \cdot \mathbf{r}+\phi_s\right)
\end{equation}
where $m^*$ is the effective mass, $\mathbf{g}_j^s$ is the \emph{s}th shell of moiré reciprocal lattice vectors and $\Delta(\mathbf{r})$ is the potential felt by holes at the valence band maximum (VBM). This potential attract holes at the AB/BA stacking regions, forming an effective moiré hole lattice [Fig. 4(b)]. To investigate the effect of spatial divisibility, we define our unit cell to intersect these AB/BA regions, effectively cutting the moiré sites under open boundary conditions [Fig. 4(a,b)].

The band structure of tbWS$_2$ for a twist angle of $\theta=2^\circ$ [Fig. 4(c)] exhibits prominent moiré band gaps. To identify the bulk topology of tbWS$_2$ within the first gap (marked red in Fig. 4(c)) , we calculate the edge polarization $\mathbf{P}$ and fractional corner charge $Q$. These quantities serve as bulk topological indices for classifying HOTIs and can be evaluated for the $C_{3}$-symmetric tbWS$_2$ system through the following formula \cite{PhysRevB.99.245151}:
\begin{equation}
\mathbf{P}=\frac{2}{3}\left(\left[K_1^{(3)}\right]+2\left[K_2^{(3)}\right]\right)\left(\mathbf{a}_1+\mathbf{a}_2\right),
\label{eq8}
\end{equation}
\begin{equation}
Q=\frac{1}{3}\left[K_2^{(3)}\right] \bmod 1,
\label{eq9}
\end{equation}
where $[K_1^{(3)}]$ and $[K_2^{(3)}]$ denote the difference in the number of bands below the energy gap for $C_{3}$ symmetry with eigenvalue $1$ and $e^{i2\pi/3}$, respectively. As shown in Fig. 4(d), a non-zero fractional corner charge $Q=1/3$ emerges in the first moiré band gap across a wide range of twist angles, while the edge polarization $\mathbf{P}$ vanishes. This result directly demonstrates that cutting through the moiré hole sites induces a 2D mHOTI phase in tbWS$_2$. This finding is further supported by our analysis, which shows that the top six bands are topologically equivalent to the well-established $H_1^{(6)}$ HOTI model from a previous study \cite{PhysRevB.99.245151} (Supplementary Material \cite{supplementary}).

To confirm the mHOTI phase in tbWS$_2$, we calculate its ribbon spectrum [Fig. 4(e)]. The spectrum reveals several topological edge bands (red), including a gapped pair near the first bulk band gap, which is a characteristic signature of the HOTI phase. Additional edge bands below this gap suggest more complex topological states. Under full open boundary conditions, the energy spectrum [Fig. 4(f)] further displays three highly localized corner states (blue) within the HOTI gap, directly confirming the mHOTI phase. A control calculation confirms the system becomes trivial when the boundaries avoid the moiré sites, emphasizing the crucial role of moiré site termination (Supplementary Material \cite{supplementary}). Given that the HOTI gap in tbWS$_2$ exceeds $20~\text{meV}$ over a wide range of twist angles, this mHOTI phase should be robust and experimentally detectable by scanning tunneling microscopy measurement. Our results provide a universal framework for understanding mHOTI phases in other moiré systems, such as twisted bilayer graphene and boron nitride \cite{PhysRevLett.126.066401}.

In summary, we have introduced an intrinsic mHOTI phase in moiré materials, rooted in the spatial divisibility of moiré sites. Unlike atomic lattices, the large scale of moiré site allows its internal degrees of freedom to be partitioned by a physical boundary. As demonstrated in tbWS$_2$, our work establishes cutting of moiré lattice site as a new design principle for engineering moiré higher-order topology and provides clear guidance for experimental exploration. Given the crucial role of electron correlations in moiré systems, our findings pave the way for exploring correlated HOTIs and higher-order superconductivity in van der Waals heterostructures.

\begin{acknowledgments}
L.X. acknowledges the support by the National Key Research and Development Program of China (2021YFA1202902), the National Natural Science Foundation of China (Grant No. 62341404), Hangzhou Tsientang Education Foundation and the Max Planck Partner group programme. 
X.Z. and Z.S. acknowledge support by the National Key R\&D Program of China (No. 2022YFA1402702), and the National Natural Science Foundation of China (No. 12374292 and  T2525032). Work at SUSTech was supported by the National Key Research and Development Program of China (2024YFA1409101), National Natural Science Foundation of China (12374059), and the Shenzhen Basic Research Fund (JCYJ20240813095301003). The computer time was supported by the Center for Computational Science and Engineering of Southern University of Science and Technology and the Major Science and Technology Infrastructure Project of Material Genome Big-science Facilities Platform supported by the Municipal Development and Reform Commission of Shenzhen. Part of this work was supported by the Quantum Science Center of Guangdong-Hong Kong-Macao Greater Bay Area (Guangdong).
Z.F.W acknowledges the support by the National Natural Science Foundation of China (No. 12174369), and the Innovation Program for Quantum Science and Technology (No. 2021ZD0302800).
This work was also supported by the Cluster of Excellence ‘Advanced Imaging of Matter' (AIM), Grupos Consolidados (IT1453-22), and the Max Planck-New York City Center for Non-Equilibrium Quantum Phenomena. The Flatiron Institute is a division of the Simons Foundation.
\end{acknowledgments}

\bibliography{ref_v1}

\end{document}


\title{Supplemental Material for\\Intrinsic Moiré Higher-Order Topology Beyond Effective Moiré Lattice Models}

\author{Xianliang Zhou}
\thanks{These authors contributed equally.}
\affiliation{Key Laboratory of Artificial Structures and Quantum Control (Ministry of Education), School of Physics and Astronomy, Shanghai Jiao Tong University, Shanghai 200240, China}
\affiliation{Tsung-Dao Lee Institute, Shanghai Jiao Tong University, Shanghai, 200240, China}
\affiliation{Institute of Physics, Chinese Academy of Sciences, Beijing 100190, People’s Republic of China}

\author{Yifan Gao}
\thanks{These authors contributed equally.}
\affiliation{Department of Physics, Southern University of Science and Technology, Shenzhen, Guangdong 518055, China}

\author{Laiyuan Su}
\affiliation{Department of Physics, Southern University of Science and Technology, Shenzhen, Guangdong 518055, China}

\author{Z.~F. Wang}
\affiliation{Hefei National Research Center for Physical Sciences at the Microscale, CAS Key Laboratory of Strongly-Coupled Quantum Matter Physics, Department of Physics, University of Science and Technology of China, Hefei, Anhui 230026, China}
\affiliation{Hefei National Laboratory, University of Science and Technology of China, Hefei, Anhui 230088, China}

\author{Li Huang}
\affiliation{Department of Physics, Southern University of Science and Technology, Shenzhen, Guangdong 518055, China}
\affiliation{Quantum Science Center of Guangdong-HongKong-Macao Greater Bay Area (Guangdong), Shenzhen 518045, China}

\author{Angel Rubio}
\affiliation{Max Planck Institute for the Structure and Dynamics of Matter, Center for Free Electron Laser Science, 22761 Hamburg, Germany}
\affiliation{Initiative for Computational Catalysis (ICC) and Center for Computational Quantum Physics (CCQ), Simons Foundation Flatiron Institute, New York, NY 10010}

\author{Zhiwen Shi}
\email{zwshi@sjtu.edu.cn}
\affiliation{Key Laboratory of Artificial Structures and Quantum Control (Ministry of Education), School of Physics and Astronomy, Shanghai Jiao Tong University, Shanghai 200240, China}
\affiliation{Tsung-Dao Lee Institute, Shanghai Jiao Tong University, Shanghai, 200240, China}
\affiliation{Collaborative Innovation Center of Advanced Microstructures, Nanjing University, Nanjing 210093, China}

\author{Lede Xian}
\email{ldxian@tias.ac.cn}
\affiliation{Tsientang Institute for Advanced Study, Zhejiang 310024, China}
\affiliation{Songshan Lake Materials Laboratory, Dongguan, Guangdong 523808, China}
\affiliation{Max Planck Institute for the Structure and Dynamics of Matter, Center for Free Electron Laser Science, 22761 Hamburg, Germany}

\maketitle

\subsection{1. Trivial phase for a conventional unit cell in 1D model} 
\begin{figure}[h]
\includegraphics[width=0.4\linewidth]{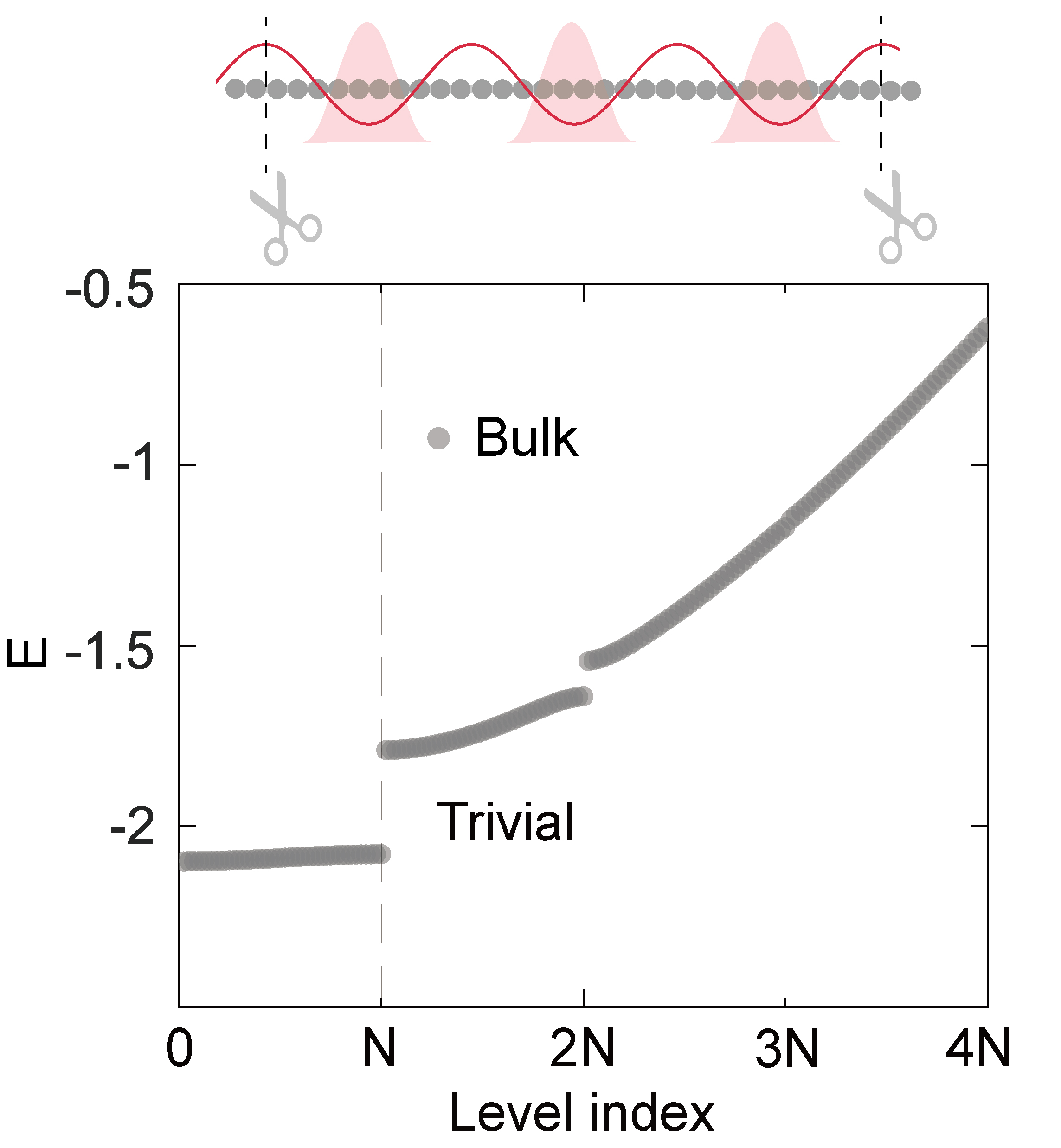}
\centering
\caption{\label{figs1} Open-boundary energy spectrum for the 1D model where the unit cell boundaries do not intersect a moiré site.}
\end{figure}

As a control calculation, we calculate the open-boundary energy spectrum of the simplified 1D model for a conventional unit cell whose boundaries aviod intersect the moiré sites. As shown in Fig. ~\ref{figs1}, while the bulk states are unaffected, boundary states are now completely absent at both the full filling of the bottom and second bands. This calculation confirms that, for a conventional unit cell definition, the system remains topologically trivial, emphasizing that cutting of moiré site is essential for inducing the topological phase.

\subsection{2. Moiré superlattice in strained bilayer graphene nanoribbon}
\begin{figure*}[h]
\includegraphics[width=0.7\linewidth]{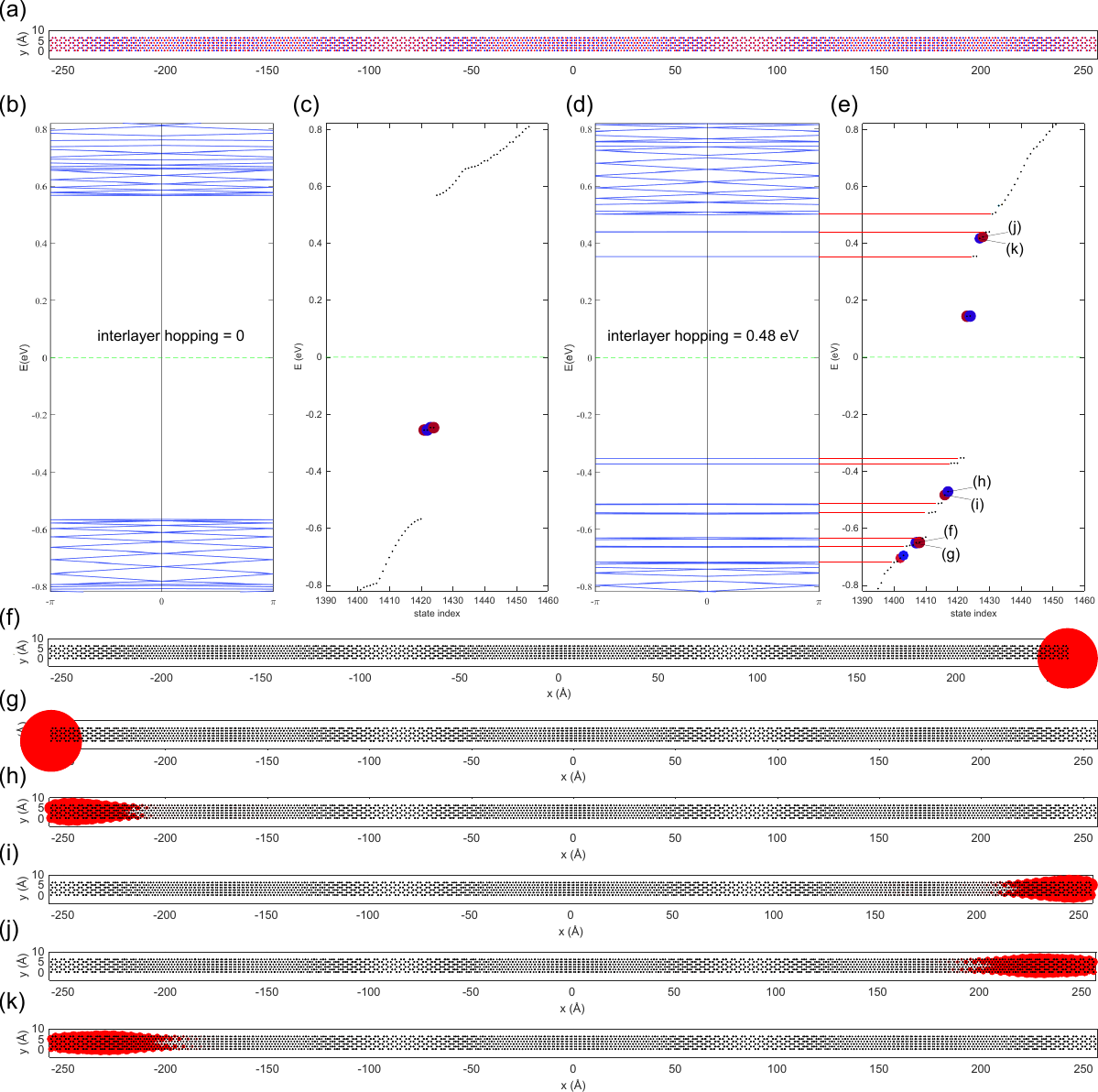}
\caption{\label{figs2} (a) A 1D moiré superlattice in bilayer graphene nanoribbon created by hetero-uniaxial strain. (b) Band structure without interlayer coupling. (c) Energy spectrum of (b). (d) Band structure with interlayer coupling. (e) Energy spectrum of (d), showing moiré-induced boundary states. (f)-(k) Wavefunctions of the boundary states marked in (e).}
\end{figure*}

A 1D graphene moiré superlattice can be realized by applying uniaxial hetero-strain to a bilayer graphene nanoribbon along the $x$-direction [Fig. ~\ref{figs2}a]. This moiré superlattice features AA-stacked regions at both ends, forming a repeating pattern with three moiré periods in the $x$-direction. To investigate the spatial divisibility of the moiré site in graphene nanoribbon, we construct a tight-binding (TB) model with the following Hamiltonian: 
\begin{equation}
    \hat{H} =\sum_{\langle i,j\rangle} t_{ij} (\hat{c}^{\dagger}_i\hat{c}_j 
    +\hat{c}^{\dagger}_j\hat{c}_i),
    \label{con:tight binding}
\end{equation}
where $\hat{c}^{\dagger}_i$ and $\hat{c}_i$ are the creation and annihilation operators for electrons in the $p_z$ orbital at site $i$. The hopping matrix element $t_{ij}$ between two $p_z$ orbitals at positions $\vec{r}_i$ and $\vec{r}_j$ is given by:
\begin{equation}
    t_{ij} =V_{pp\pi}(r)[1-(\frac{\vec{r}_{ij}\cdot\hat{z}}{r})^2]
    +V_{pp\sigma}(r)(\frac{\vec{r}_{ij}\cdot\hat{z}}{r})^2,
    \label{con:tight binding1}
\end{equation}
where $\hat{z}$ is the unit vector along the $z$-direction, and $r=|\vec{r}_i-\vec{r}_j|$ is the distance between two orbitals. The parameters $V_{pp\pi}(r)$ and $V_{pp\sigma}(r)$ are defined as $V_{pp\pi}(r) = \gamma_0e^{(a_0-r)/r_0}$ and $V_{pp\sigma}(r)=\gamma_1e^{(d_0-r)/r_0}$ , with $a_0=1.42~\mathrm{\AA}$, $d_0=3.35~\mathrm{\AA}$, $r_0=0.451~\mathrm{\AA}$, $\gamma_0=-2.7~\text{eV}$ and $\gamma_1=0.48~\text{eV}$ \cite{PhysRevB.86.125413}.

Without interlayer coupling, our calculations show intrinsic graphene-zigzag boundary states in the middle of the energy gap [Fig. ~\ref{figs2}(c)] \cite{PhysRevB.54.17954}. When interlayer coupling is introduced, a periodic moiré potential emerges, opening a series of moiré band gaps [Fig. ~\ref{figs2}(d)]. New boundary states appear within these gaps due to the cutting of moiré sites [Fig. ~\ref{figs2}(e)]. The wavefunctions of these boundary states are shown in Fig. ~\ref{figs2}(f)–(k).

\begin{figure}[h]
\centering
\includegraphics[width=0.7\linewidth]{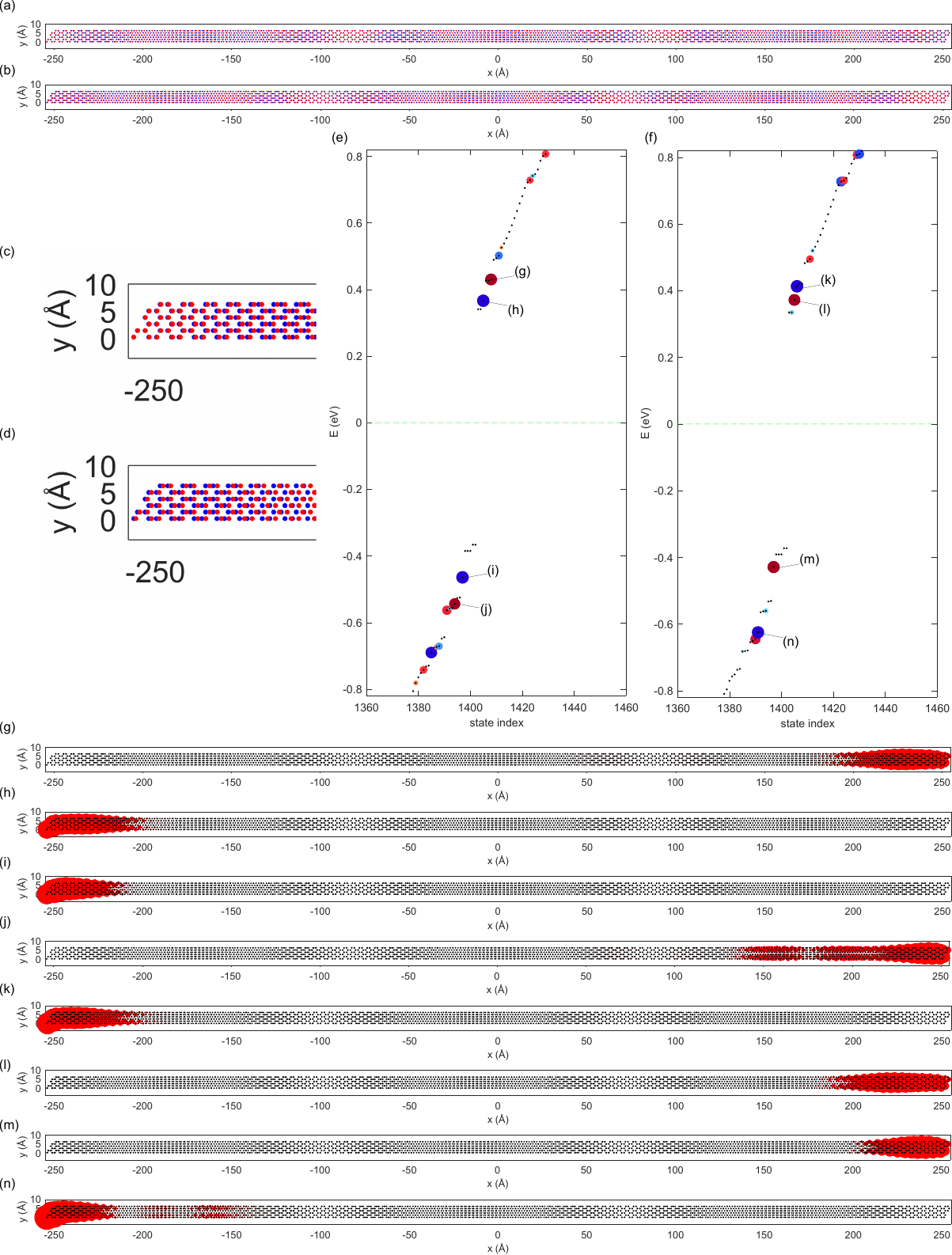}
\caption{\label{figs3} (a) Bilayer graphene nanoribbon superlattice with a diagonal boundary configuration, making both edges armchair-type. (b) Same as (a) but with a sliding structure where the edges are not perfectly aligned with the AA-stacked regions. (c-d) Zoomed-in views of the boundaries in (a) and (b), respectively. (e-f) Energy spectra for the structures shown in (a) and (b). (g-n) Wavefunctions of the boundary states marked in (e) and (f).}
\end{figure}

To distinguish these newly induced states from the intrinsic graphene-zigzag boundary states, we introduce a diagonal cut at the boundaries, making both edges armchair-type [Fig. ~\ref{figs3}(a)]. Our results reveal that these moiré-induced boundary states persist even when both edges are of the armchair type, as illustrated in Fig. ~\ref{figs3}(e). Remarkably, these boundary states remain robust even when the boundaries are not perfectly aligned with the AA-stacked regions, as shown in the sliding structures in Fig. ~\ref{figs3}(b) and (d), where the edges shift from the exact AA-stacked configuration.

\subsection{3. DFT-simulated Beryllium-Hydrogen atomic chain}
\begin{figure}[h]
\includegraphics[width=0.7\linewidth]{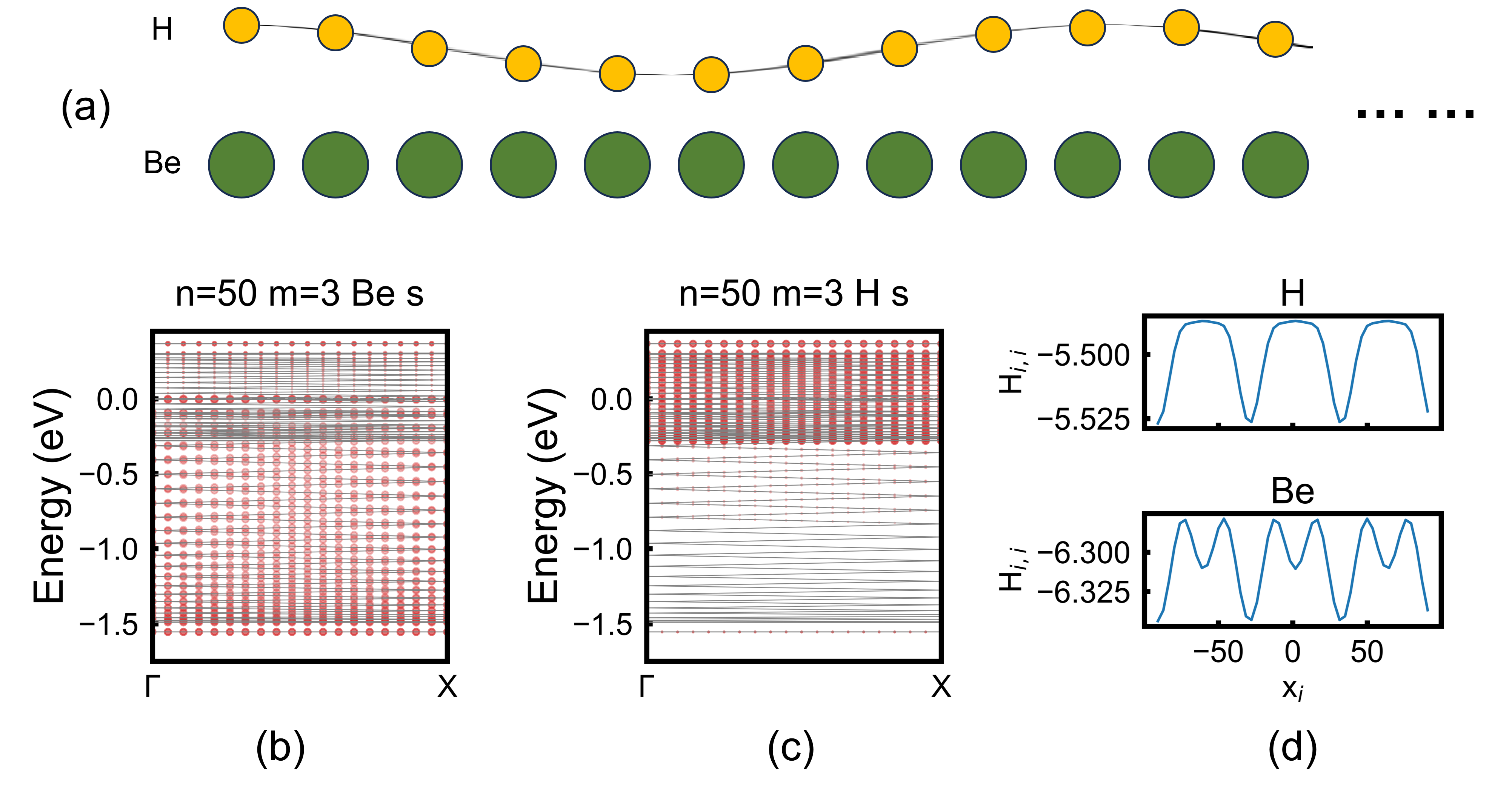}
\centering
\caption{\label{fig.dftmodel}(a) Schematic of the 1D atomic chain. (b)(c) DFT calculated band structure of the periodic atomic chain, where red circles shows the orbital contribution of Be s and H s orbitals, respectively. (d) The on-site energy of Be and H sites fitted with wannier function.}
\end{figure}

In addition to the bilayer graphene nanoribbon, we constructed a 1D atomic chain composed of Beryllium-Hydrogen (Be-H) pairs, as shown in Fig. ~\ref{fig.dftmodel}(a). We control the chain's moiré potential by varying the distance between Be and H atoms, $d_{Be-H}$, while keeping the distance between neighboring Be and H atoms fixed $d_{Be-Be}=d_{H-H}$=3.5 {\AA}. The distance $d_{Be-H}$ is defined by a periodic function: $d_{Be-H}=d_0 + v_{max}f(\frac{2im\pi}{n})$. Here, $v_{max}$=0.5 {\AA} is the effective amplitude of the potential, $d_0$=3 {\AA} is the equilibrium distance, and $n$ is the total number of Be/H atoms. The moiré period is determined by $m$, which is the periodicity of the function $f(x)\in\{\cos(x), -\cos(x),\sin(x)\}$ and corresponds to the number of "moiré atoms" or moiré lattice sites in the system.

First-principles calculations are performed within the density functional theory (DFT) framework using the projected augmented-wave (PAW) method \cite{PhysRevB.50.17953} as implemented in the Vienna ab initio simulation package (VASP) \cite{KRESSE199615,PhysRevB.54.11169}. The exchange-correlation term is treated in the generalized gradient approximation (GGA) of Perdew-Burke-Ernzerhof (PBE) \cite{PhysRevLett.77.3865}. The Kohn-Sham orbitals are expanded in a plane wave basis set with an energy cutoff of 600 eV. Visualization of the geometric structures and real-space wave function illustrations is performed using the VESTA package \cite{Momma:db5098}. To eliminate artificial interactions between periodic neighboring slabs, a vacuum space of more than 20 {\AA} is introduced along the directions perpendicular to the 1D chain.

The DFT-calculated band structures for $n=50$, $m=3$ and $f=\cos(x)$ are plotted in Fig. ~\ref{fig.dftmodel}(b) and (c). The lower 50 bands originate from Be $s$ orbitals, while the upper 50 are primarily contributed by H atoms. Due to strong band mixing near the Fermi level, we focus our discussion only on the topmost and bottom bands, where the identification of boundary states is clearer. We also calculated the on-site energy, which is the moiré potential, of each atomic site by fitting to a set of Wannier orbital bases. As displayed in Fig. ~\ref{fig.dftmodel}(d), the on-site potential increases as $d_{Be-H}$ decreases. The concave region in the Be curve is attributed to the van der Waals potential becoming critical as the Be-H distance shrinks. To simulate open boundary conditions that spatially divide the moiré lattice sites, we introduced a vacuum layer at the cell boundary. This approach allows us to investigate different chain terminations by modifying the periodic function $f$.

\subsection{{f(x)=cos(x)}}
When $f(x)=\cos(x)$, the cutting edge corresponds to the lowest point of the periodic potential. To effectively determine the location of the boundary states, we calculate two key quantities for each eigenstate $\psi_n$: the inverse participation ratio (IPR) and the wavefunction center $r_c$. The IPR quantifies the degree of localization, with larger values indicating stronger localization. The wavefunction center, $r_c$, indicates where the state is spatially located. These are given by the following equations:

\begin{equation}
    IPR(\psi_n)=\frac{\sum^N_{i=1}{|\psi_n(\vec{r_i})|^4}}{(\sum^N_{i=1}{|\psi_n(\vec{r_i})|^2})^2}
    \label{ipr}
\end{equation}
\begin{equation}
    r_c = \frac{\sum^N_{i=1}{r_i|\psi_n(\vec{r_i})|^2}}{\sum^N_{i=1}{|\psi_n(\vec{r_i})|^2}}, r_i\in [-0.5,0.5]
    \label{wavefunction_center}
\end{equation}

\begin{figure*}[h]
\includegraphics[width=0.7\linewidth]{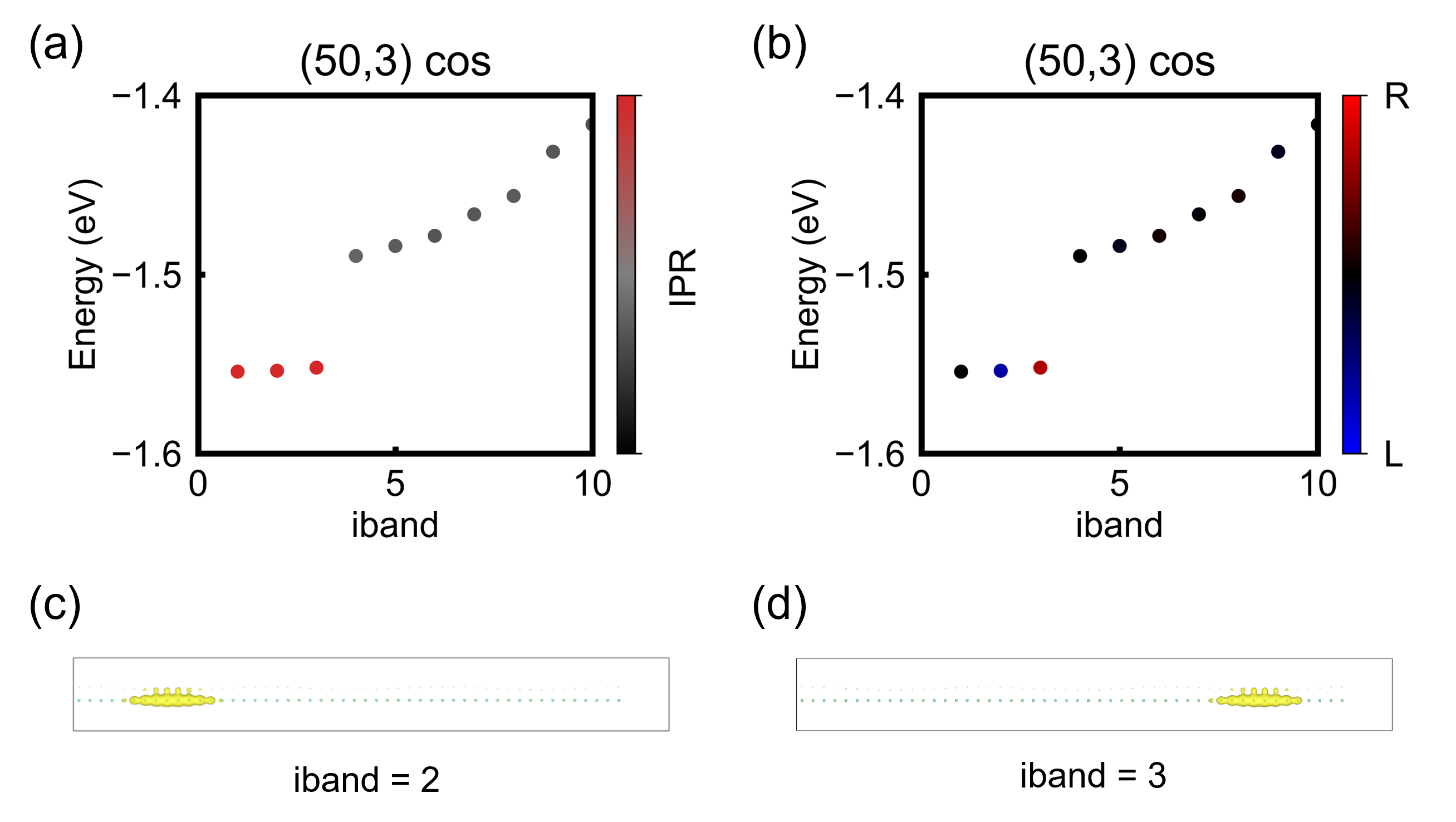}
\centering
\caption{\label{fig.cos} (a-b) IPR and $r_c$ of eigenstates when f(x)=cos(x). (c-d) Real space illustration of the squared wave functions $|\psi_2|^2$ and $|\psi_3|^2$, respectively.}
\end{figure*}

As shown in Fig.~\ref{fig.cos}(a-b), the first three eigenstates exhibit strong localization. However, only the eigenstates in bands $n=2$ and $n=3$ correspond to boundary states. No boundary states are observed in the upper region of the bands, which belongs to the Be-H hybridized area.

\begin{figure*}[h]
\includegraphics[width=0.7\linewidth]{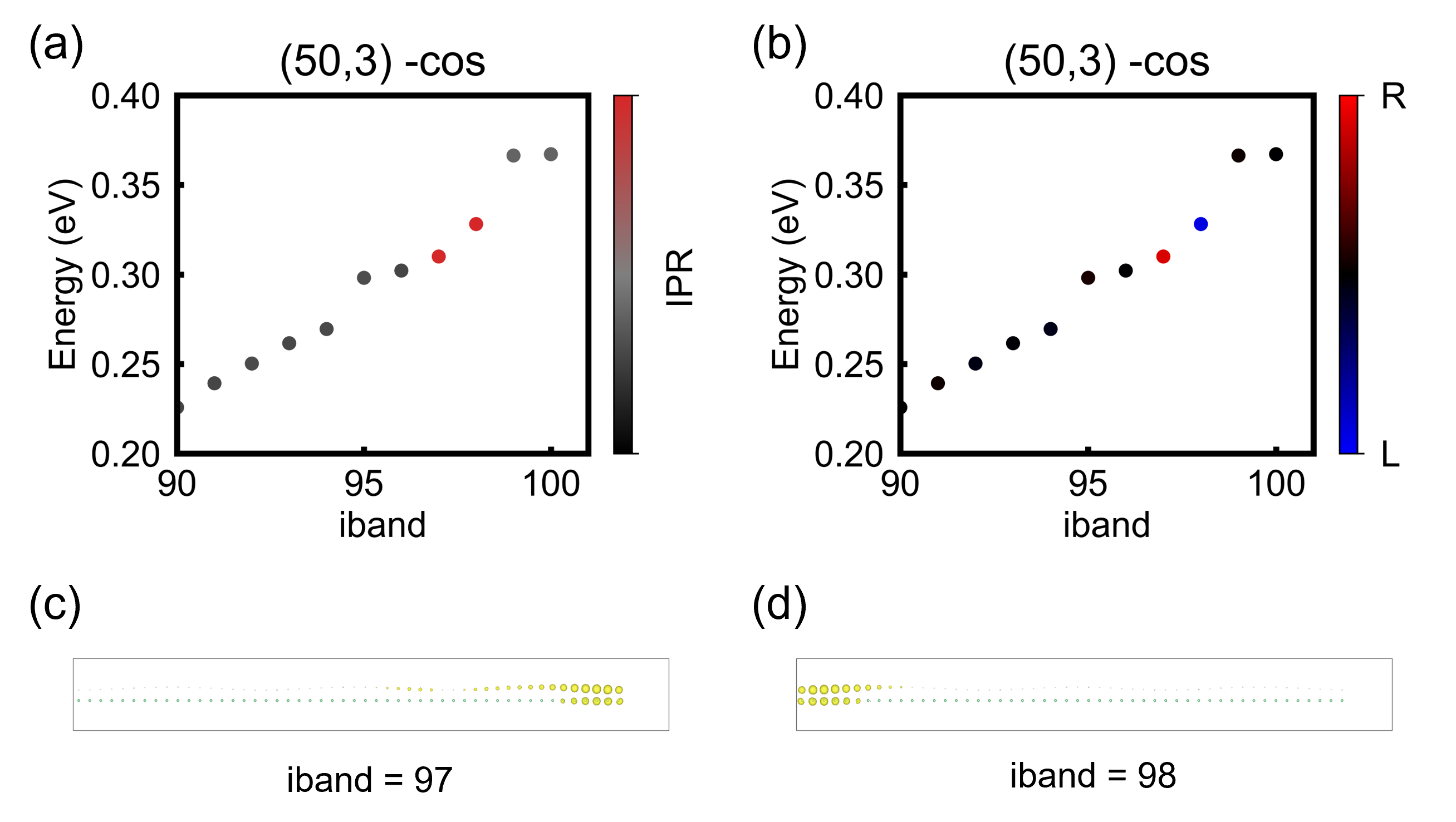}
\centering
\caption{\label{fig.mcos} (a-b) IPR and $r_c$ of eigenstates when f(x)=-cos(x). (c-d) Real space illustration of the squared wave functions $|\psi_{97}|^2$ and $|\psi_{98}|^2$, respectively.}
\end{figure*}

\subsection{{f(x)=-cos(x)}}
When $f(x)=-\cos(x)$, the cutting edge corresponds to the highest point of the periodic potential for H atoms. This configuration facilitates the identification of boundary states in the upper region of the bands, as depicted in Fig.~\ref{fig.mcos}. For Be atoms, the cutting edge is located in the concave valley of the potential, but remains significantly higher than the potential minimum. Consequently, it is challenging to identify boundary states in the lower region of the bands. We propose that the boundary states in the Be region might reside within the mixed area near the Fermi level.

\subsection{{f(x)=sin(x)}}
\begin{figure*}[h]
\includegraphics[width=0.7\linewidth]{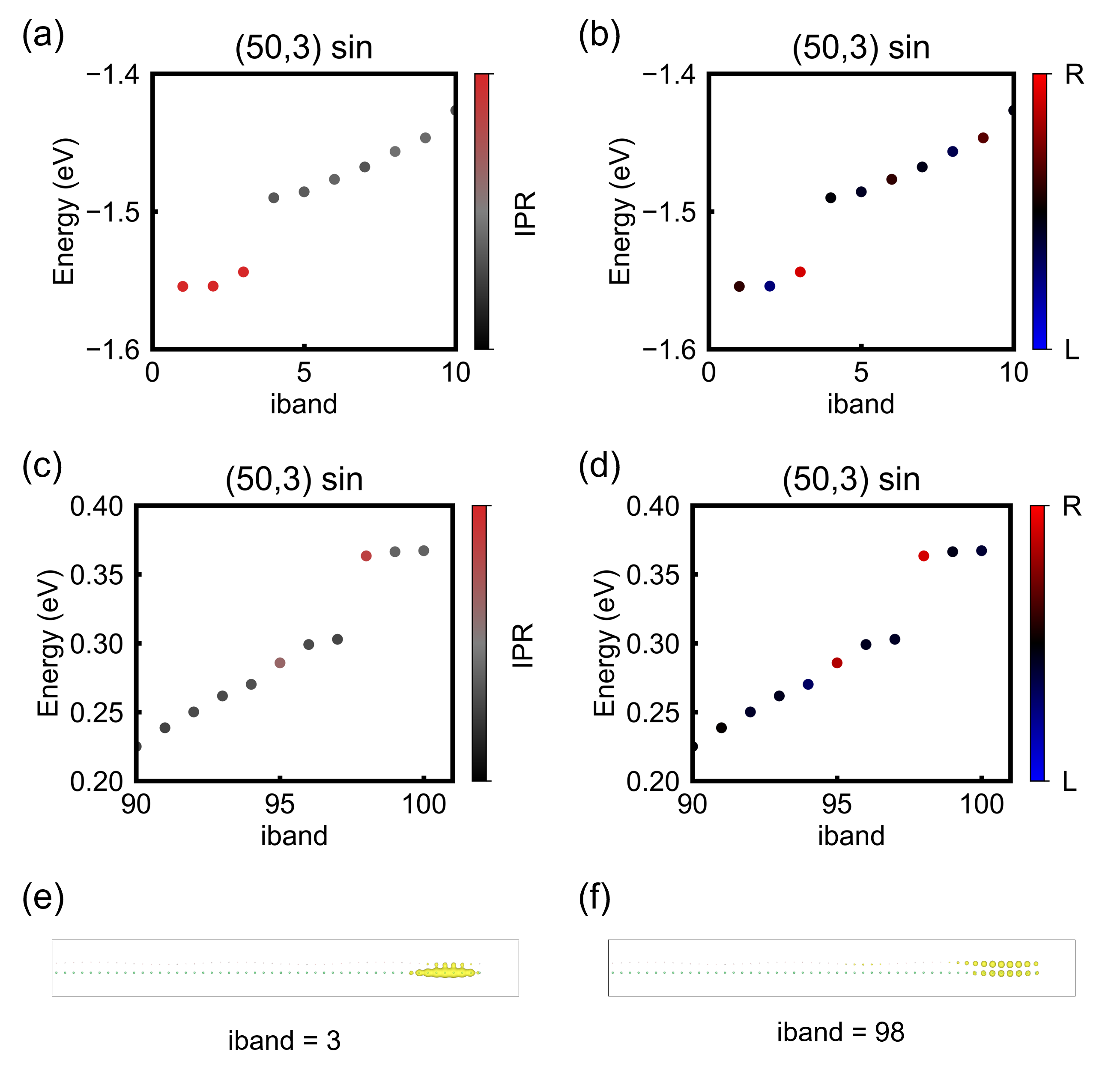}
\centering
\caption{\label{fig.sin} (a-b) IPR and $r_c$ of eigenstates when f(x)=sin(x). (c-d) Real space illustration of the squared wave functions $|\psi_{3}|^2$ and $|\psi_{98}|^2$, respectively.}
\end{figure*}

When $f(x)=\sin(x)$, the cutting edge is positioned in the intermediate region between the high and low valleys of the periodic potential. In this configuration, only one boundary state is observed in both the upper and lower regions of the bands, as illustrated in Fig.~\ref{fig.sin}.

~\\
~\\

\subsection{4. Trivial phase for a conventional unit cell in 2D model}
\begin{figure*}[h]
\includegraphics[width=0.7\linewidth]{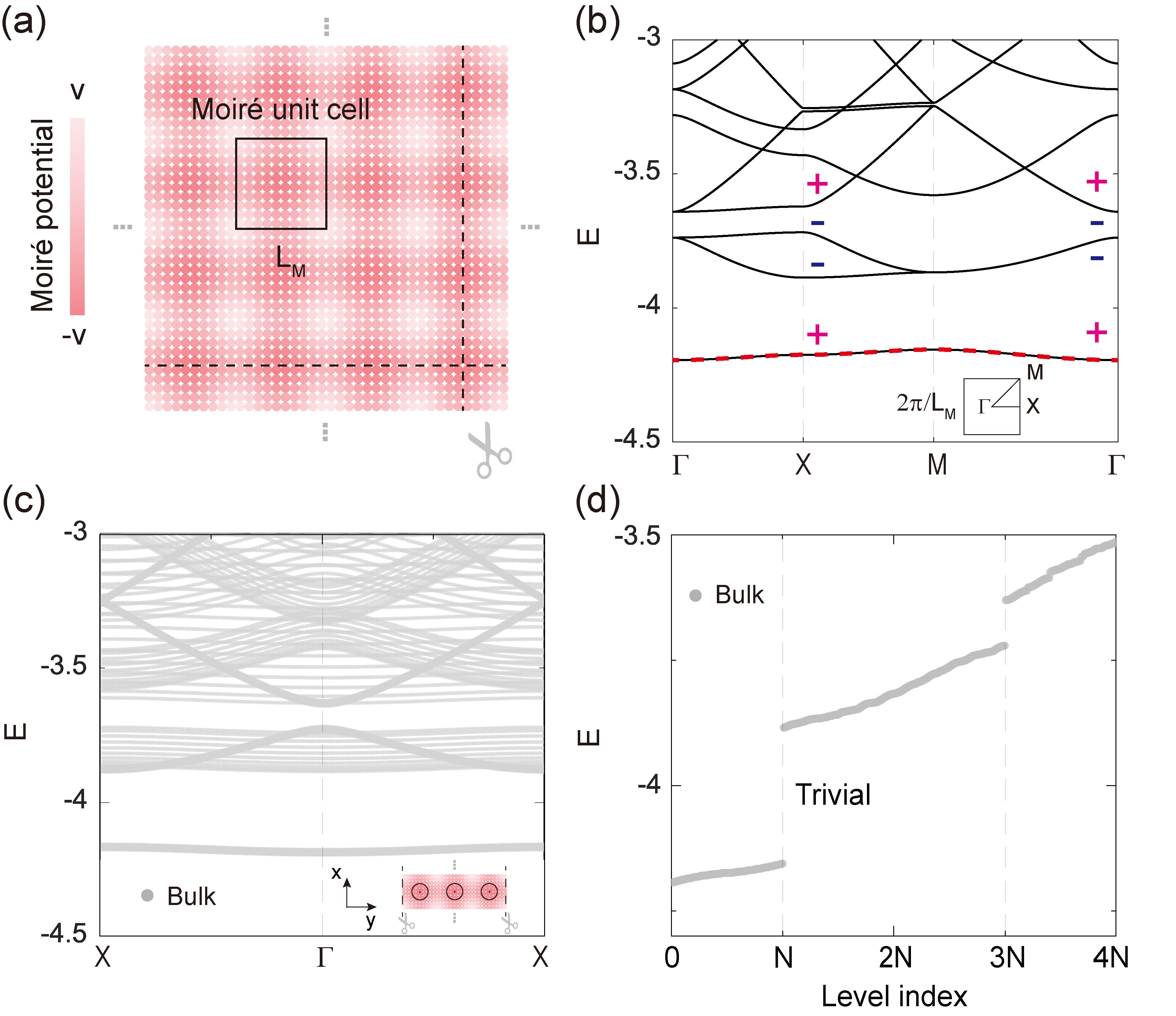}
\centering
\caption{\label{figs8}Simplified 2D model with a conventional unit cell. (a) Schematic of the simplified model. The solid box marks the moiré unit cell. (b) The moiré band structure for $v/t=0.3$ and $L_M=10$. The red dashed line represents a fit of effective lattice model. Parity eigenvalues at high-symmetry points are labeled by “$\pm$”. (c) Ribbon spectrum of the band structure in (b), finite along the $y$-direction ($N_y=10$). Inset shows the ribbon geometry. (d) Energy spectrum of the band structure in (b), of a finite flake with $N=N_x\times N_y=10\times 10=100$ unit cells.}
\end{figure*}

Fig.~\ref{figs8}(b) shows the moiré band structure for the simplified 2D model system using a conventional unit cell whose boundaries avoid the moiré sites [Fig.~\ref{figs8}(a)]. Similar to the 1D case in the main text, the electronic states of the lowest moiré band are localized, forming an effective square lattice. The low-energy physics of this effective lattice is described by the effective Hamiltonian:
\begin{equation}
H= \sum_{\mathbf{R}, \mathbf{R}^{\prime}} t\left(\mathbf{R}-\mathbf{R}^{\prime}\right) c_{\mathbf{R} }^{\dagger} c_{\mathbf{R}^{\prime}},
\label{effective}
\end{equation}
where $\mathbf{R}$ denotes the moiré square lattice sites.

This effective lattice model accurately reproduces the band dispersion of the lowest band in the 2D system [Fig.~\ref{figs8}(b)]. Although this dispersion is identical to the topological case shown in the main text, the parity eigenvalues of the four lowest bands are different at the $X$ point. As a result, these bands have a trivial polarization ($\mathbf{P}=0$) and topology.

The ribbon spectrum [Fig. ~\ref{figs8}(c)] confirms the absence of topological edge states. Similarly, the spectrum for a finite flake with full open boundaries [Fig. ~\ref{figs8}(d)] shows no in-gap corner states. These results for a conventional unit cell highlight that the topological boundary states discussed in the main text are a direct consequence of cutting of the moiré sites.

\subsection{5. Robustness to Boundary Shifts}
Because each moiré site is spatially extended over tens of atoms, the topological boundary states arising from cutting are robust against small shifts in the boundary's position, which act as a weak local perturbation.

\begin{figure}[h]
\includegraphics[width=0.7\linewidth]{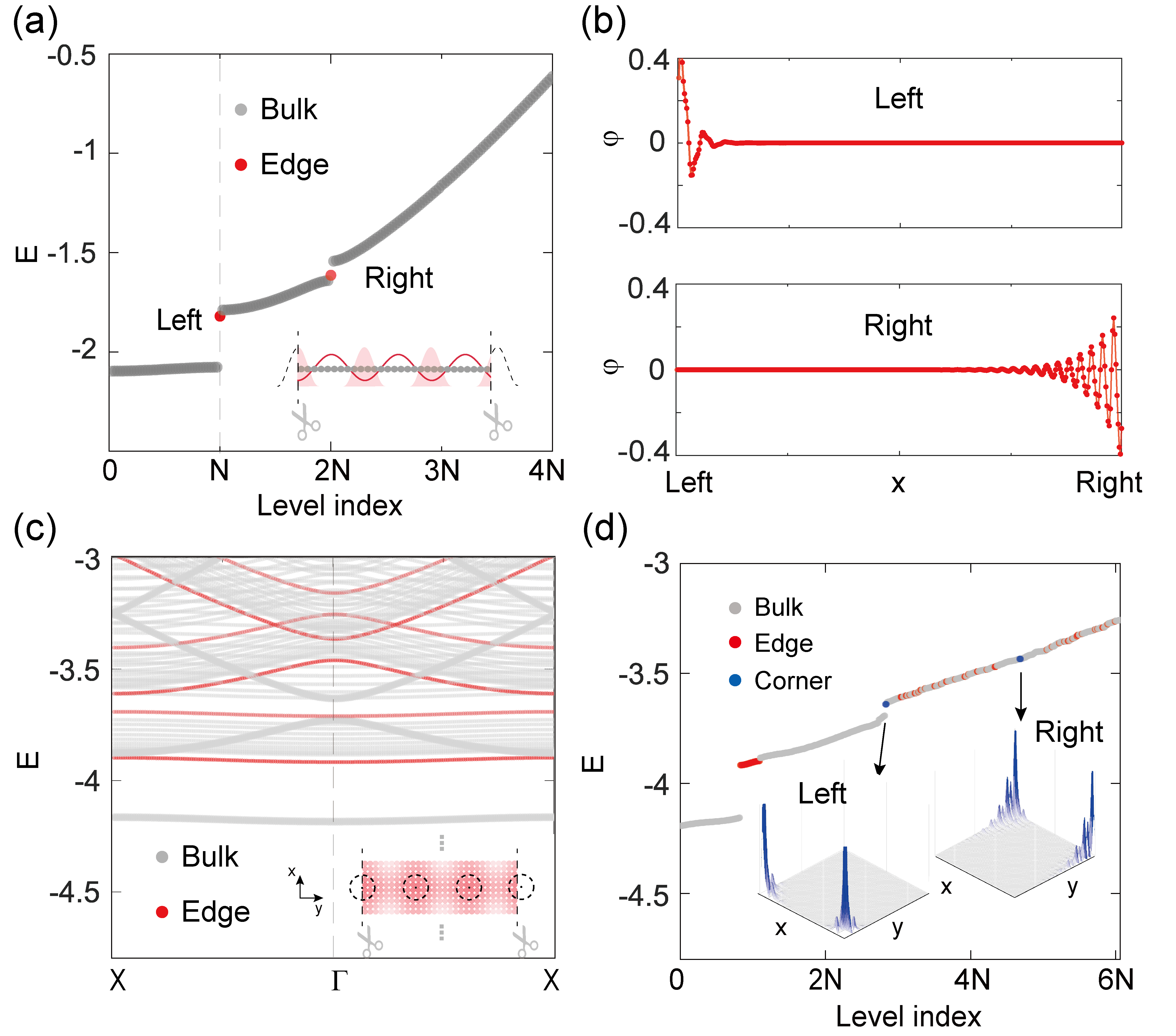}
\centering
\caption{\label{perturbation}Robustness of boundary states against boundary shifts. (a) Energy spectrum of the 1D model with a shifted right boundary (see inset).  (b) Wavefunctions of two boundary states in (a). (c) Ribbon spectrum of the 2D model with similarly shifted boundaries. (d) Energy spectrum of (c) with inset showing the charge density of the corner states.}
\end{figure}

To test the robustness of the boundary states, we shifted the right boundary of the simplified 1D model by one atomic site [inset, Fig. \ref{perturbation}(a)]. The resulting energy spectrum [Fig. \ref{perturbation}(a)] shows that while the boundary states persist within the gap, this asymmetric perturbation lifts their degeneracy. This energy splitting is reflected in their spatial distribution [Fig. \ref{perturbation}(b)]: the unperturbed left-edge state is unaffected, while the state at the shifted right edge moves to a higher energy and becomes more delocalized. Further shifts would cause this state to merge within the bulk spectrum, demonstrating the robustness of the topological phase up to a critical perturbation strength.

Figure \ref{perturbation}(c) and (d) show the ribbon and finite flake spectra for the simplified 2D model with a similarly shifted boundary. The results are analogous to the 1D case: the topological boundary states are robust against this local perturbation, a tolerance that simplifies experimental implementation. Moreover, the fact that the boundary states' energy and spatial distribution can be modified by such shifts highlights a novel approach for engineering their properties in moiré superlattices.

~\\
~\\
~\\

\subsection{6. Topological Equivalence of tbWS$_2$ to the $H_1^{(6)}$ model}
We find that the highest six energy bands of our twisted bilayer Tungsten disulfide (tbWS$_2$) calculation in the main text are topologically equivalent to those of the well-established $H_1^{(6)}$ higher-order topological insulator (HOTI) model \cite{PhysRevB.99.245151}. This model describes a HOTI on a hexagonal lattice with the Hamiltonian:
\begin{equation}
H_1^{(6)}(\mathbf{k})=\left(\begin{array}{cccccc}
0 & t_0 & e^{i \mathbf{k} \cdot \mathbf{a}_2} & 0 & e^{-i \mathbf{k} \cdot \mathbf{a}_3} & t_0 \\
t_0 & 0 & t_0 & e^{-i \mathbf{k} \cdot \mathbf{a}_3} & 0 & e^{-i \mathbf{k} \cdot \mathbf{a}_1} \\
e^{-i \mathbf{k} \cdot \mathbf{a}_2} & t_0 & 0 & t_0 & e^{-i \mathbf{k} \cdot \mathbf{a}_1} & 0 \\
0 & e^{i \mathbf{k} \cdot \mathbf{a}_3} & t_0 & 0 & t_0 & e^{-i \mathbf{k} \cdot \mathbf{a}_2} \\
e^{i \mathbf{k} \cdot \mathbf{a}_3} & 0 & e^{i \mathbf{k} \cdot \mathbf{a}_1} & t_0 & 0 & t_0 \\
t_0 & e^{i \mathbf{k} \cdot \mathbf{a}_1} & 0 & e^{i \mathbf{k} \cdot \mathbf{a}_2} & t_0 & 0
\end{array}\right),
\end{equation}
where $t_0$ is the intra-cell hopping, and $\mathbf{a}_1=(1,0), \mathbf{a}_2=\left(\frac{1}{2}, \frac{\sqrt{3}}{2}\right),\mathbf{a}_3=\left(\frac{1}{2}, -\frac{\sqrt{3}}{2}\right)$ are the lattice vectors.

\begin{figure}[h]
\includegraphics[width=0.7\linewidth]{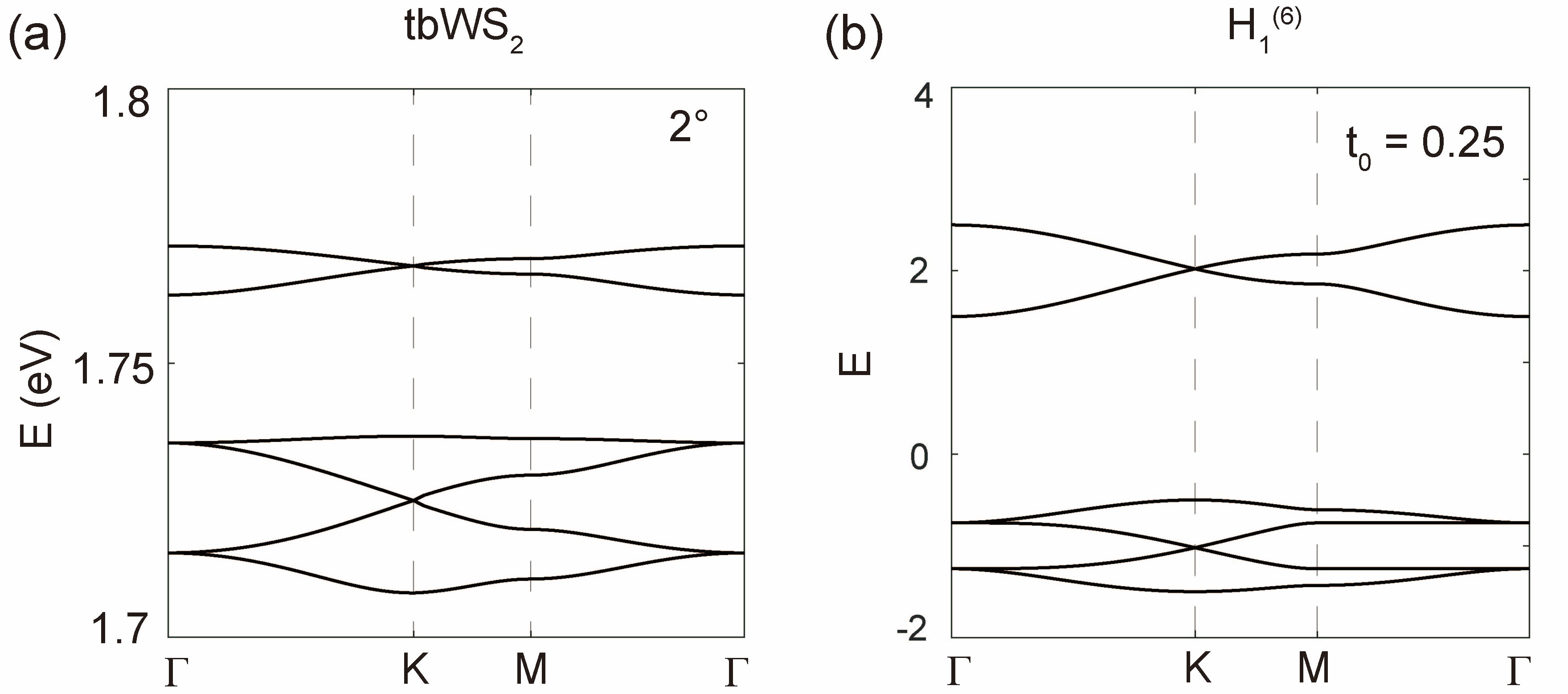}
\centering
\caption{\label{H16}(a) Highest six energy bands of tbWS$_2$. (b) Band structure of $H_1^{(6)}$ model for $t_0=0.25$.}
\end{figure}

The band structure of this Hamiltonian, calculated for $t_0=0.25$, is shown in Fig. \ref{H16}(b) and is remarkably similar to the six highest bands of tbWS$_2$ [Fig. 4(c) in the main text]. The $H_1^{(6)}$ model at 2/3-filling is a known HOTI with topological indices $\chi^{(6)}=\left(\left[M_1^{(2)}\right],\left[K_1^{(3)}\right]\right)=(0,2)$ \cite{PhysRevB.99.245151}. We find that the corresponding bands in our tbWS$_2$ calculation share the same rotation eigenvalues as this model, yielding identical topological indices. This topological equivalence provides further proof that cutting of the moiré sites drives tbWS$_2$ into a HOTI phase.

\subsection{7. Control Calculation: Trivial Phase in tbWS$_2$}
\begin{figure}[h]
\includegraphics[width=0.7\linewidth]{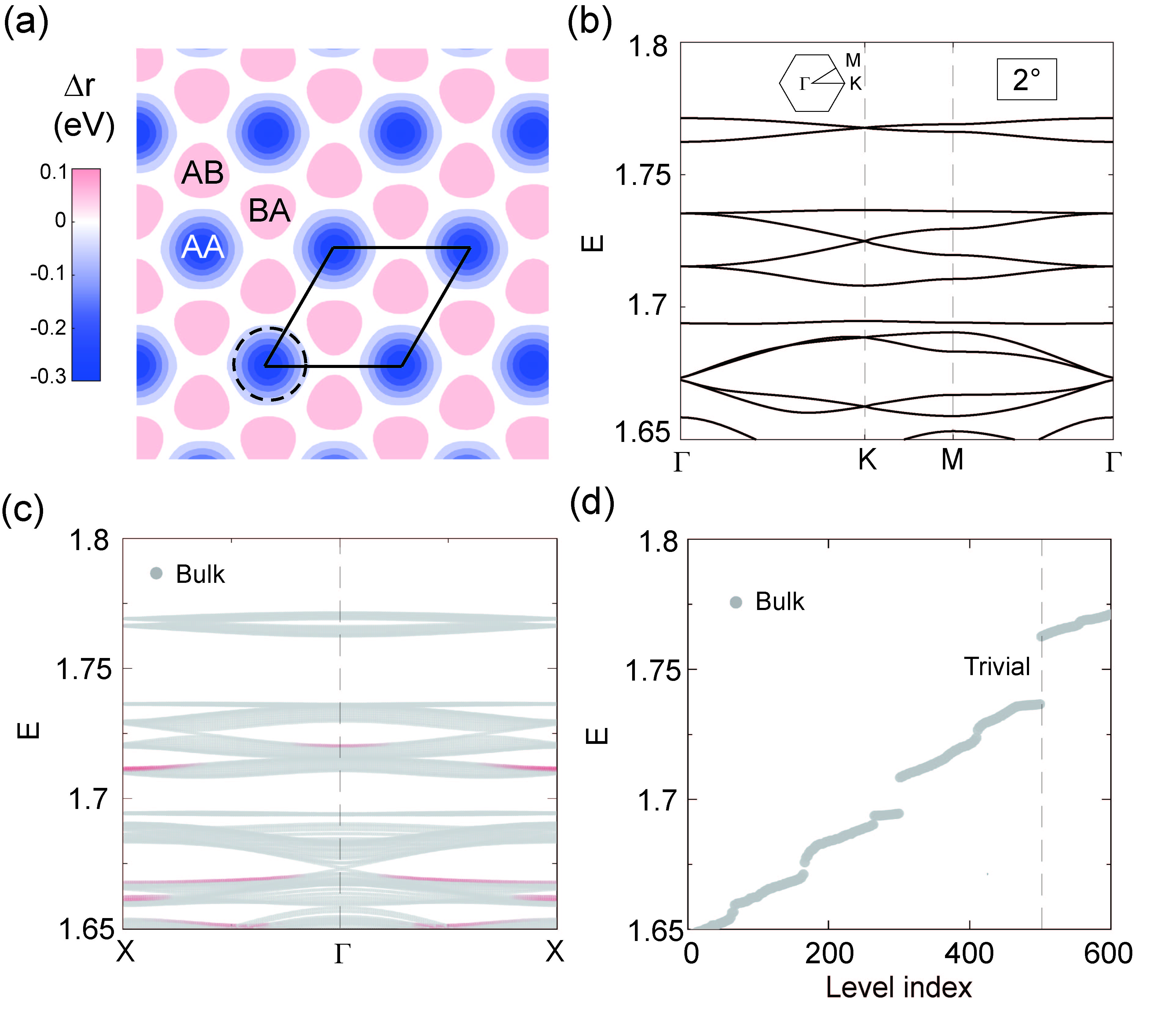}
\centering
\caption{\label{tbWS2}Control calculation showing the trivial phase in tbWS$_2$ for a conventional unit cell. (a) Moiré potential felt by holes and moiré unit cell  whose boundaries avoid the potential maxima (moiré hole sites). (b) The resulting bulk band structure. (c) The corresponding ribbon spectrum. (d) Energy spectrum under full open-boundary conditions for a $10\times10$ supercell.}
\end{figure}

As a control calculation, we investigate a tbWS$_2$ configuration with a conventional unit cell whose boundaries avoid the moiré hole sites [Fig. \ref{tbWS2}(a)]. The calculated ribbon [Fig. \ref{tbWS2}(c)] and finite flake [Fig. \ref{tbWS2}(d)] spectra both show a complete absence of the topological boundary states seen in the main text. The disappearance of boundary states confirms that the tbWS$_2$ system remains in a topological trivial phase when the moiré sites are not cut, emphasizing that cutting of moiré sites is the essential mechanism for realizing the HOTI phase in moiré systems.

\bibliography{ref_sup}